\documentclass[%
 reprint,
 amsmath,amssymb,
 aps,
]{revtex4-2}

\usepackage{graphicx}
\usepackage{dcolumn}
\usepackage{bm}
\usepackage{hyperref}

\begin{document}

\preprint{APS/123-QED}

\title{How well can superconducting nanowire single-photon detectors resolve photon number?}

\author{Timon Schapeler}
\email{timon.schapeler@upb.de}
\author{Niklas Lamberty}
\author{Thomas Hummel}
\author{Fabian Schlue}
\author{Michael Stefszky}
\author{Benjamin Brecht}
\author{Christine Silberhorn}
\author{Tim J. Bartley}
\affiliation{Institute for Photonic Quantum Systems, Department of Physics, Paderborn University, Warburger Str. 100, 33098 Paderborn, Germany}

\date{\today}

\begin{abstract}
We apply principal component analysis (PCA) to a set of electrical output signals from a commercially available superconducting nanowire single-photon detector (SNSPD) to investigate their photon-number-resolving capability. We find that the rising edge as well as the amplitude of the electrical signal have the most dependence on photon number. Accurately measuring the rising edge while simultaneously measuring the voltage of the pulse amplitude maximizes the photon-number resolution of SNSPDs. Using an optimal basis of principle components, we show unambiguous discrimination between one- and two-photon events, as well as partial resolution up to five photons. This expands the use-case of SNSPDs to photon-counting experiments, without the need of detector multiplexing architectures. 
\end{abstract}

\maketitle


\section{Introduction}
The ability to count photons is desirable in a great many quantum optics experiments and applications, not least quantum computing paradigms~\cite{kok2007linear}, quantum communication~\cite{gisin2002quantum,gisin2007quantum} and multiphoton metrology~\cite{slussarenko2017unconditional}. While some energy-resolving detectors function at the single-photon level, they are often slow devices with limited timing resolution~\cite{calkins2011faster}. Instead, single photon detectors, which detect the presence or absence of photons, but cannot resolve photon number, can offer quasi-photon number resolution when configured in multiplexing architectures~\cite{paul1996photon,achilles2003fiber,fitch2003photon,banaszek2003photon,dauler2007multi,cheng2023a}. Within this class of detectors, superconducting nanowire single-photon detectors (SNSPDs) have become the gold standard due their unmatched single-to-noise ratio~\cite{chiles2022new} and high timing resolution~\cite{korzh2020demonstration}.
 
While long regarded as a threshold detector, careful analysis of the detector signal has recently demonstrated that they also show limited photon-number information in their electrical output signal. 
The first evidence of photon-number resolution using an SNSPD was presented by Cahall et al.~\cite{cahall2017multi}, which was attributed to a time- and photon-number dependent hotspot resistance of the nanowire. By investigating the different rise times from the detector's electrical output signal, they show evidence up to four-photon detection events. 
These findings are consistent with a generalized electro-thermal model~\cite{nicolich2019universal} based on Ref.~\cite{kerman2009electrothermal}. This model links hotspot growth and the rising edge of the detector's electrical output signal.
Another approach to gain photon-number resolution from SNSPDs was presented by Zhu et al.~\cite{zhu2020resolving}, where the amplitude of the detector output signal directly correlates to photon number up to five photons. This was achieved by combining a single superconducting nanowire with an impedance-matching taper, which made the amplitude of the output signal sensitive to the number of hotspots induced by absorbed photons, which was predicted by Bell et al. in 2007~\cite{bell2007photon}.
Further, in an attempt to be more resistant against noise, Endo et al. used an oscilloscope to record reference waveforms corresponding to different photon numbers, which could then be used to distinguish different events by waveform pattern matching. They found that using a subset of the rising edge of the SNSPD output trace improves the photon-number discrimination for larger input states, compared to merely using a single point in time~\cite{endo2021quantum}.
Alternative approaches for photon-number resolution involve analyzing the rising edge of the output signal. One method utilizes a linear fit to a section of the rising edge, with photon number determined from the corresponding slope measured using an oscilloscope~\cite{sempere-llagostera2022reducing}. Another method measures the variation in slew-rate of the SNSPD output signal, employing a constant-threshold time tagger to determine the difference in arrival time of detection events relative to a trigger signal~\cite{davis2022improved}.

The most recent method to resolve photon number with a single SNSPD builds on this idea and measures the relative time difference between a trigger signal and the rising and falling edges of the electrical detector output signal, which enhances the discrimination of different photon numbers~\cite{sauer2023resolving}.

While investigating the full electrical output signal of a detector maximizes the amount of information from a detection event, this is not practical in experiments where photon-number information must be extracted in real time. The question naturally arises: which aspects of the electrical output provide most information about the photon number?
To answer this question, in this paper we use a popular technique from multivariate statistics known as principal component analysis (PCA)~\cite{abdi2010principal} to deduce maximum information about photon number from a minimum number of data points.

\begin{figure*}[ht] 
\includegraphics{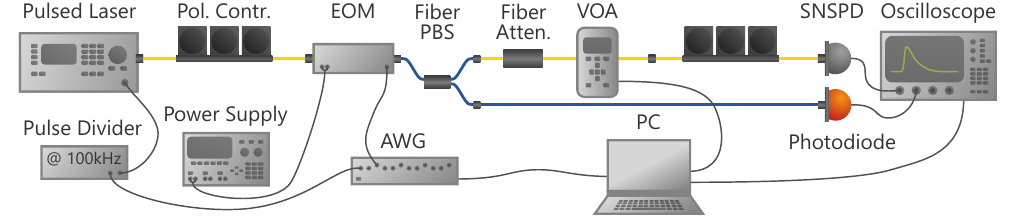}
\caption{We use a femtosecond-pulsed laser at $80~\mathrm{MHz}$, which is pulse-picked down to $100~\mathrm{kHz}$ using an electro-optic modulator (EOM). The EOM receives a square-wave function from an arbitrary waveform generator (AWG) and a DC offset voltage from a power supply. The polarization controller and the fiber polarizing beam splitter (PBS) are used to optimize the optical power incident on the SNSPD. A fiber attenuator and a variable optical attenuator are used to attenuate the laser light and set different mean photon numbers. The second polarization controller is used to optimize the signal on the SNSPD. The second port of the fiber PBS is connected to a fast photodide, which is used as the trigger for the oscilloscope. The yellow and blue fibers represent single-mode and polarization-maintaining fiber respectively.}
\label{fig:setup}
\end{figure*}

In general, PCA is a dimensionality-reduction technique that determines a set of optimal orthogonal basis functions (principal components) describing the maximum amount of information in a data set. That means it transforms the data set, containing lots of variables, into a smaller data set, being less accurate but with reduced complexity. PCA has already been applied to the electrical output signals from transition-edge sensors (TESs)~\cite{humphreys2015tomography}, where it was successfully used to reduce the complexity of the data set by focusing on the first principal components. 

In this work, we make use of the full electrical output signal of the commercially available SNSPD, maximizing the amount of information from any detection event. From this we will use PCA to investigate the extent to which SNSPDs can resolve photon number, and to identify which features of the electrical output signal are most relevant for maximizing the photon-number-resolving capability. Following that, we make use of the knowledge from the principal component analysis and use a time tagger to verify our findings. We use a more application-oriented method of recording the relative time differences between a trigger signal and the rising and falling edges of the SNSPD electrical output signal~\cite{sauer2023resolving}.

\section{Experimental Setup}
The experiment consists of recording the electrical output signal from a commercially available SNSPD with an oscilloscope. The experimental setup can be seen in Fig.~\ref{fig:setup}. We use a $<200~\mathrm{fs}$ pulsed laser, as recent theoretical simulations suggest that longer delays between absorbed photons decreases resolvability~\cite{dryazov2023modeling}. Since the SNSPD cannot be operated at the repetition rate of the pulsed laser of $80~\mathrm{MHz}$, we pick every 800th pulse to reduce the repetition rate to $100~\mathrm{kHz}$. This is done by dividing the synchronisation output of the laser with a pulse divider, which is then used as a trigger for an arbitrary waveform generator (AWG) to generate a $8.3~\mathrm{ns}$ pulse for the pulse-picking electro-optic modulator (EOM). We use a power supply to set a DC offset voltage to the EOM, in order to optimize the extinction ratio of the pulse-picking system. The polarization controller before, and the fiber polarizing beam splitter (PBS) after the EOM are used to optimize the polarization of the light towards the SNSPD. One port of the PBS is connected to a fiber attenuator (in order to attenuate the light close to the single-photon level) and two variable optical attenuators (VOA) control the precise mean photon number impinging on the detector. The second polarization controller is used to optimize the detection efficiency of the polarization-dependent SNSPD. The second port of the fiber PBS is connected to a fast photodiode ($1.2~\mathrm{GHz}$ bandwidth), which acts as a trigger for the oscilloscope. We use a $21~\mathrm{GHz}$ bandwidth oscilloscope with $128~\mathrm{GSa/s}$ to record the electrical signal of the SNSPD. This gives us a sampling period of roughly $8~\mathrm{ps}$, while maintaining the full information of the electrical signal of the detector. The SNSPD has a timing jitter of $11.3~\mathrm{ps}$, specified by the manufacturer.

The oscilloscope is operated in a sequential acquisition mode, where 1000 electrical output traces are recorded at once and then streamed to a computer. One trace consists of $3\times10^4$ points, which corresponds to a length in time of approximately $234~\mathrm{ns}$. We scan trough mean photon numbers from $0.5$ to $5$ in steps of $0.5$ and record $1\times10^5$ traces per mean photon number. An exception is the mean photon number of $0.5$ we record twice as many traces, since a lot of pulses contain zero photons. Every instance of zero impinging photons results in an empty trace, which contains no relevant information for the PCA. We randomize the order of the 1100 measurements, to avoid any bias towards photon number that might occur due to possible drifts during the measurement. 

In a second measurement we substitute the oscilloscope for a time tagger with a time resolution of $<1.9~\mathrm{ps}$. This measurement is used to confirm our findings of the subsequent principal component analysis method, in order to maximize the photon-number-resolving capability of the commercial SNSPD using a time tagger.

The mean photon numbers as a function of attenuation are calibrated before the measurement. This is done by recording the count rate ($\mathrm{CR}$) of the SNSPD over the full operating range (from dark counts until saturation, where the count rate equals the repetition rate ($\mathrm{RR}$) of the experiment) with a time tagger, by sweeping the attenuation using the VOAs. Assuming Poissonian statistics of the laser, the mean detected photon number per pulse is then given by
\begin{equation}
    \bar{n}=-\mathrm{ln}\left(1-\frac{\mathrm{CR}}{\mathrm{RR}}\right)\,.
\end{equation}

\begin{figure}[ht]
\includegraphics{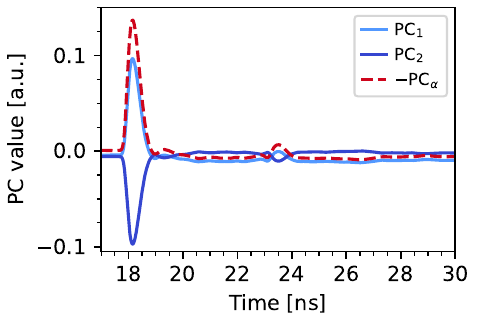}
\caption{First two principal components (basis functions) $\mathrm{PC}_1$, $\mathrm{PC}_2$ and the principal component that results from the optimal angle projection $-\mathrm{PC}_{\alpha}$ of the two-dimensional histogram in Fig.~\ref{fig:pca_2dhist}(a). Only the relevant time regime is shown, the basis functions are flat otherwise.}
\label{fig:principal_components}
\end{figure}

\section{Results}
The full data set of electrical output signals consists of $1.1\times10^6$ traces (also called samples) and $3\times10^4$ points per trace, i.e., time-dependent voltages (also called features). However, we only use $1.9\times10^4$ points as this is enough to contain the complete SNSPD electrical trace. In a first step we filter out traces from sub-optimal pulse-picking (i.e., traces with multiple peaks or traces at the wrong time delay with respect to the trigger signal) as well as empty traces (when no photon was detected, after the trigger signal from the fast photodiode). This means that in the PCA we will not investigate the zero-photon (vacuum) component of the coherent state, as we want to investigate the photon-number-resolving capability of the detector. Since the discrimination of click and no-click events is trivial, we focus only on the click events (at least one photon). 

\begin{figure*}
\includegraphics{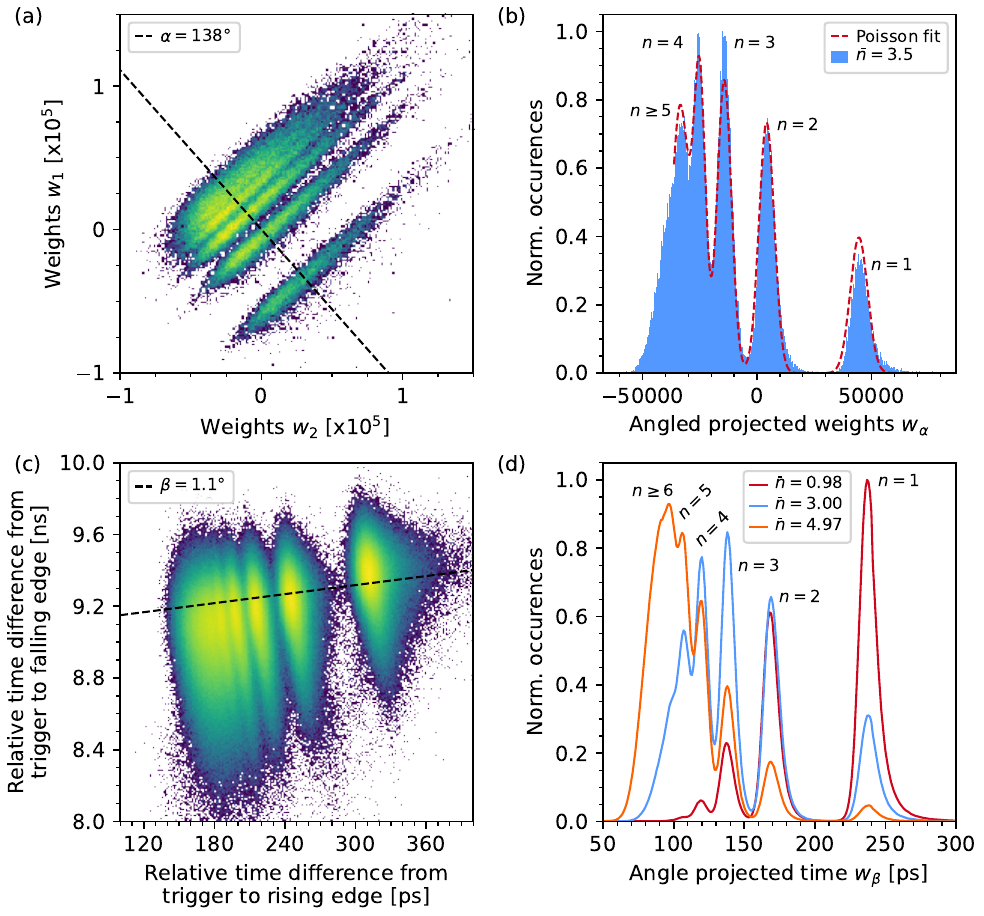}
\caption{(a) Two-dimensional histogram of the weights of the first two principal components from the PCA model for a mean photon number of $\bar{n}=3.5$. These two components show a photon-number dependency, which can be seen by the clear differentiation between the ellipses. For smaller mean photon numbers, the occupation of the ellipses shifts to the lower right and for higher mean photon numbers to the upper left. (b) Histogram of the optimal projection at an angle of $\alpha=138^{\circ}$ and a multi-Gaussian fit with an enforced Poisson distribution (red dashed line). (c) Two-dimensional histogram of the relative time difference between trigger signal and rising and falling edge. The full data set is displayed, which is the sum of the individual measurements for the three mean photon numbers indicated in (d). (d) Optimal projection at an angle of $\beta=1.1^{\circ}$ of the two-dimensional histogram of (c) for three photon numbers (as labeled in the legend).}
\label{fig:pca_2dhist}
\end{figure*}

As a next step, we fit the PCA model with a subset of 1000 samples from each mean photon number (using more samples did not change the fit). This yields the principal components $\mathrm{PC}_1$ and $\mathrm{PC}_2$ in Fig.~\ref{fig:principal_components}, which describe the most information of the data set. The model is then successively applied to the full data set in order to find the weights corresponding to the principal components for each sample. From our analysis we find that only the first two principal components show photon-number dependency. Plotting the weights of the first component against the weights of the second component in a two-dimensional histogram shows a clear differentiation between photon numbers (see Fig.~\ref{fig:pca_2dhist}(a), where the mean photon number is $\bar{n}=3.5$).
\begin{figure*}[th]
\includegraphics{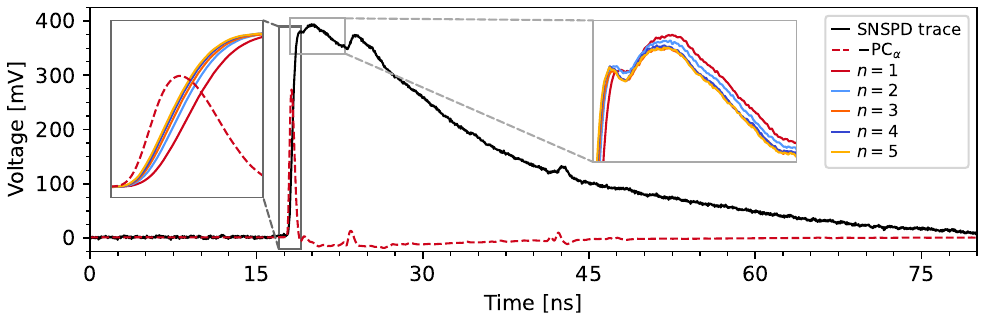}
\caption{Electrical output signal of the SNSPD, as well as the angled principal component $-\mathrm{PC}_{\alpha}$ (scaled by a factor of 2000 for better comparison). The left inset shows rising edges of the traces for different photon numbers, where an earlier rising edge correlates with a higher photon number. The right inset shows the peaks of the traces, where a reduced height correlates with a higher photon number. Colors correspond to different photon numbers, as labeled in the legend. For reduced influence of electrical noise an average over 20 traces is shown for both insets.}
\label{fig:traces}
\end{figure*}

Due to the angle of the ellipses, the components on their own will not reveal the optimal photon-number-resolving capability of the detector. However, projecting the histogram on an angle of $\alpha=138^\circ$ (as indicated by the black dashed line in Fig.~\ref{fig:pca_2dhist}(a)), the maximal photon-number resolution can be unveiled. This projection is required as the principal components describe the maximal variance in the data set which can be a mixture from various information sources. The projection distills the maximal photon-number resolution from the principal components. This leads to the histogram shown in Fig.~\ref{fig:pca_2dhist}(b) for a mean photon number of $\bar{n}=3.5$, where the angle projected weights are calculated as $w_{\alpha}=w_2\mathrm{sin}(\alpha)+w_1\mathrm{cos}(\alpha)$. After the angled projection only one basis function will describe the photon-number dependency. Interpreting this basis function reveals which features of the electrical output trace of an SNSPD need to be observed when it comes to photon-number resolution. We find that the relative time difference between the trigger signal and the rising edge (peak of $-\mathrm{PC}_{\alpha}$ around $18~\mathrm{ns}$ in Fig.~\ref{fig:principal_components}), as well as the peak amplitude of the electrical output signal from the SNSPD (part of $-\mathrm{PC}_{\alpha}$ after the peak) contain the most photon-number information.

In a recent study, the relative time difference between a trigger signal and the rising edge as well as the falling edge was observed in a two-dimensional histogram using a time tagger~\cite{sauer2023resolving}. In order to compare our oscilloscope measurement (utilizing PCA to show us the best method to retrieve photon-number information) with the method from Ref.~\cite{sauer2023resolving}, we also recorded a two-dimensional histogram of rising and falling edge using a time tagger with a time resolution of $<1.9~\mathrm{ps}$ on the same commercial SNSPD. In the end this needs less data analysis to show photon-number resolution and is thus more application-oriented. We use the same experimental setup shown in Fig.~\ref{fig:setup}, except with a time tagger that records the relative time differences from trigger signal to the rising and falling edge of the trace. The photon-number resolution depends on the voltage threshold of the time tagger. We find that at half peak height (corresponding to $185~\mathrm{mV}$) the separation between photon numbers is best. Afterwards, we plot the complete data set recorded with the time tagger in a two-dimensional histogram (see Fig.~\ref{fig:pca_2dhist}(c)), find the optimal projection angle ($\beta=1.1^{\circ}$), and plot the distributions for different mean photon numbers (see Fig.~\ref{fig:pca_2dhist}(d)). Comparing the resulting distributions in Fig.~\ref{fig:pca_2dhist}(b) and Fig.~\ref{fig:pca_2dhist}(d) reveals a very similar performance. 

In order to visualize why the two methods lead to similar results, we show an electrical output signal of the SNSPD in Fig.~\ref{fig:traces} and highlight the key features for photon-number resolution as identified by the PCA. In the insets of Fig.~\ref{fig:traces}, we show an average over 20 SNSPD traces for different photon-number events as identified by the PCA. An increased photon number correlates with a reduced time difference between a trigger signal and the rising edge. This is consistent with previous studies where the rising edge was observed in order to show photon-number resolution using an SNSPD~\cite{cahall2017multi,nicolich2019universal,endo2021quantum,sempere-llagostera2022reducing,davis2022improved,sauer2023resolving}. 

Additionally, we see that an increased photon number also correlates with a reduced peak amplitude. This appears to contradict earlier work which showed a slight increase in peak amplitude~\cite{cahall2017multi}. However, their experiment included readout electronics with a high bandwidth amplifier, which essentially converts high-speed events to higher amplitudes. In contrast, due to our cabling and amplifiers with limited bandwidth, our readout electronics effectively contains an additional low-pass filter, penalizing higher-speed features in the trace. 

Thus, utilizing the falling edge relative to a trigger signal to increase the photon-number resolution is also consistent with our findings. A reduced pulse amplitude correlates with a shorter relative time difference between a trigger signal and the falling edge of the SNSPD electrical output trace at a constant threshold (see right inset in Fig.~\ref{fig:traces}). While it is certainly challenging to make simultaneous precise time and voltage measurements, this can be circumvented by only focusing on the time measurement of rising and falling edges using a precise time tagger. As mentioned above, the threshold of the time tagger needs to be chosen carefully in order to maximize photon-number resolution. The left inset in Fig.~\ref{fig:traces} shows that the separation between different photon-number events appears largest at half peak height, which is consistent with the optimized threshold set for the time tagger measurement.

Comparing the optimal projection for the PCA in Fig.~\ref{fig:pca_2dhist}(a) and the time tagger analysis in Fig.~\ref{fig:pca_2dhist}(c) shows very close agreement. Due to the increased time resolution of the time tagger, this method is suitable to circumvent the need to measure the peak amplitude of the electrical output signal to optimize photon-number resolution.
\begin{figure}[t]
\includegraphics{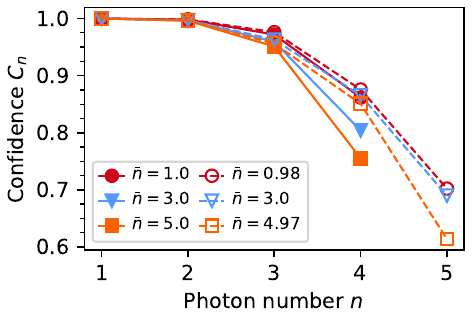}
\caption{Confidence $C_n$ vs. photon number for the analysis of the oscilloscope data using PCA (solid symbols) and the time tagger data (hollow symbols) for three mean photon numbers.}
\label{fig:confidence}
\end{figure}
In order to characterize the photon-number resolution of the SNSPD we make use of the confidence measure
\begin{equation}\label{eqn:con}
    C_n=\int_{-\infty}^\infty \frac{p(w|n)^2p(n)}{p(w)}\mathrm{d}w \,,
\end{equation}
which was proposed by Humphreys et al.~\cite{humphreys2015tomography}. This measure describes the probability that a certain detector response yields the correct photon number. Here $p(n)$ is the probability of an $n$-photon input, which is in our case a Poissonian distribution. $p(w)$ is the overall probability distribution of the angle projected weight $w_{\alpha}$ (time $w_{\beta}$) of the two-dimensional histograms in Fig.~\ref{fig:pca_2dhist}(b) and Fig.~\ref{fig:pca_2dhist}(d). At last, $p(w|n)$ is the probability $p(w)$ given an incident photon number $n$. This is derived from the Gaussian functions fitted to the histograms shown in Fig.~\ref{fig:pca_2dhist}(b) and Fig.~\ref{fig:pca_2dhist}(d). 

In Fig.~\ref{fig:confidence} we show the confidence $C_n$ for the two data sets (PCA and time tagger method) for three mean photon numbers. Since this work is focused on the read out of the SNSPD, we do not account for the detector efficiency of $(77\pm3)\%$, which means the confidence only takes into account the correct read out of the photon number absorbed by the SNSPD and not the actual incident photon number. 
As can be seen in Fig.~\ref{fig:pca_2dhist}(b) for the PCA data set and in Fig.~\ref{fig:pca_2dhist}(d) for the time tagger data set, photon-number contributions overlap significantly for $n\geq5$ and $n\geq6$, respectively. Therefore we only calculate the confidence up to $n=4$ (PCA) and $n=5$ (time tagger). Figure~\ref{fig:confidence} shows that the confidence decreases with increasing photon number, which results from the fact that the spacing of the ellipses (corresponding to photon-number) in Fig.~\ref{fig:pca_2dhist}(a) and Fig.~\ref{fig:pca_2dhist}(c) decreases.

The confidence for the time tagger measurement is always slightly higher, which can be explained by the higher timing resolution of $<1.9~\mathrm{ps}$ compared to the timing resolution of the oscilloscope of roughly $8~\mathrm{ps}$. That means that focusing merely on the temporal measurement of the rising and falling edges with a time tagger can effectively substitute measuring the SNSPD signal amplitude (proposed by the PCA). This makes a time tagger the ideal candidate to extract photon-number information from SNSPDs. Additionally, the width of the ellipses of the time tagger measurement in Fig.~\ref{fig:pca_2dhist}(c) is limited by the system jitter of the experimental setup. Reducing the system jitter should therefore improve the photon-number discrimination, more noticeably for higher photon numbers.

\section{Conclusion}
Using principal component analysis (a tool from multivariate statistics) on a set of electrical output signals from an SNSPD allows us to investigate which features of the SNSPD output trace are most relevant for photon-number resolution. Our analysis indicates that the rising edge, as well as the amplitude of the electrical output signal show the most photon-number information. Thus, simultaneously measuring the rising edge and the amplitude of the SNSPD output signal should maximize the photon-number resolution. We additionally show that a more application-oriented approach of recording the relative time difference from a trigger signal to the rising and falling edge of the SNSPD output signal in a two-dimensional histogram (recently shown in Ref.~\cite{sauer2023resolving}) can effectively substitute the voltage measurement by a temporal measurement of the falling edge of the output signal using a time tagger. We explain this by the much better timing resolution of $<1.9~\mathrm{ps}$ of the time tagger compared to the timing resolution of the oscilloscope. We also characterize the photon-number resolution by calculating a confidence measure~\cite{humphreys2015tomography} (which is the probability that a certain detector response yields the correct photon number) of the two methods presented in this work. We find that the confidence using the time tagger method is always slightly higher and that the time tagger can in principle resolve higher photon numbers, due to the better timing resolution. This makes the time tagger an excellent tool to enable photon-number resolution with SNSPDs.

\begin{acknowledgments}
The authors thank Klaus Jöns for lending the oscilloscope and Vladyslav Dyachuk for fruitful discussions. Partially funded by the European Union (ERC, QuESADILLA, 101042399). Views and opinions expressed are however those of the author(s) only and do not necessarily reflect those of the European Union or the European Research Council Executive Agency. Neither the European Union nor the granting authority can be held responsible for them. This work has received funding from the German Ministry of Education and Research within the PhoQuant project (grant number 13N16103).
\end{acknowledgments}


\bibliography{references}

\end{document}